\input harvmac.tex
\input amssym.tex
\catcode`\@11\relax
\newif\ify@autoscale \y@autoscaletrue \def\Yautoscale#1{\ifnum #1=0
  \y@autoscalefalse\else\y@autoscaletrue\fi}
\newdimen\y@b@xdim
\newdimen\y@boxdim \y@boxdim=13pt
\def\Yboxdim#1{\y@autoscalefalse\y@boxdim=#1}
\newdimen\y@linethick    \y@linethick=.3pt
\def\Ylinethick#1{\y@linethick=#1}
\newskip\y@interspace \y@interspace=0ex plus 0.3ex
\def\Yinterspace#1{\y@interspace=#1}
\newif\ify@vcenter   \y@vcenterfalse
\def\Yvcentermath#1{\ifnum #1=0 \y@vcenterfalse\else\y@vcentertrue\fi}
\newif\ify@stdtext   \y@stdtextfalse
\def\Ystdtext#1{\ifnum #1=0 \y@stdtextfalse\else\y@stdtexttrue\fi}
\newif\ify@enable@skew   \y@enable@skewfalse
\expandafter\ifx\csname enableskew\endcsname\relax
 \y@enable@skewfalse \else \y@enable@skewtrue\fi
\def\y@vr{\vrule height0.8\y@b@xdim width\y@linethick depth 0.2\y@b@xdim}
\def\y@emptybox{\y@vr\hbox to \y@b@xdim{\hfil}}
\ify@enable@skew
 \def\y@abcbox#1{\if :#1\else
   \y@vr\hbox to \y@b@xdim{\hfil#1\hfil}\fi}
 \def\y@mathabcbox#1{\if :#1\else
   \y@vr\hbox to \y@b@xdim{\hfil$#1$\hfil}\fi}
\else
 \def\y@abcbox#1{\y@vr\hbox to \y@b@xdim{\hfil#1\hfil}}
 \def\y@mathabcbox#1{\y@vr\hbox to \y@b@xdim{\hfil$#1$\hfil}}
\fi
\def\y@setdim{%
  \ify@autoscale%
   \ifvoid1\else\typeout{Package youngtab: box1 not free! Expect an
     error!}\fi%
   \setbox1=\hbox{A}\y@b@xdim=1.6\ht1 \setbox1=\hbox{}\box1%
  \else\y@b@xdim=\y@boxdim \advance\y@b@xdim by -2\y@linethick
  \fi}
\newcount\y@counter
\newif\ify@islastarg
\def\y@lastargtest#1,#2 {\if\space #2 \y@islastargtrue
  \else\y@islastargfalse\fi}
\def\y@emptyboxes#1{\y@counter=#1\loop\ifnum\y@counter>0
  \advance\y@counter by -1 \y@emptybox\repeat}
\def\y@nelineemptyboxes#1{%
  \vbox{%
    \hrule height\y@linethick%
    \hbox{\y@emptyboxes{#1}\y@vr}
    \hrule height\y@linethick}\vskip-\y@linethick}
\def\yng(#1){%
  \y@setdim%
  \hskip\y@interspace%
  \ifmmode\ify@vcenter\vcenter\fi\fi{%
  \y@lastargtest#1,
  \vbox{\offinterlineskip
    \ify@islastarg
     \y@nelineemptyboxes{#1}
    \else
     \y@ungempty(#1)
    \fi}}\hskip\y@interspace}
\def\y@ungempty(#1,#2){%
  \y@nelineemptyboxes{#1}
  \y@lastargtest#2,
  \ify@islastarg
   \y@nelineemptyboxes{#2}
  \else
   \y@ungempty(#2)
  \fi}
\def\y@nelettertest#1#2. {\if\space #2 \y@islastargtrue
  \else\y@islastargfalse\fi}
\def\y@abcboxes#1#2.{%
  \ify@stdtext\y@abcbox#1\else\y@mathabcbox#1\fi%
  \y@nelettertest #2.
  \ify@islastarg\unskip%
   \ify@stdtext\y@abcbox{#2}\else\y@mathabcbox{#2}\fi%
  \else\y@abcboxes#2.\fi}
 \newdimen\y@full@b@xdim
 \newcount\y@m@veright@cnt
\ify@enable@skew
 \def\y@get@m@veright@cnt#1#2.{%
   \if :#1 \advance\y@m@veright@cnt by 1\y@get@m@veright@cnt#2.\fi}
 \let\y@setdim@=\y@setdim
 \def\y@setdim{%
   \y@setdim@ \y@full@b@xdim=\y@b@xdim
   \advance\y@full@b@xdim by 1\y@linethick}
 \def\y@m@veright@ifskew#1{
   \y@m@veright@cnt=0 \y@get@m@veright@cnt#1.
   \moveright \y@m@veright@cnt\y@full@b@xdim}
\else
 \def\y@m@veright@ifskew#1{}
\fi
\def\y@nelineabcboxes#1{%
  \y@nelettertest #1.
  \ify@islastarg
   \y@m@veright@ifskew{#1}
    \vbox{
      \hrule height\y@linethick%
      \hbox{\ify@stdtext\y@abcbox#1\else\y@mathabcbox#1\fi\y@vr}
      \hrule height\y@linethick}\vskip-\y@linethick
  \else
   \y@m@veright@ifskew{#1}
    \vbox{
      \hrule height\y@linethick%
      \hbox{\y@abcboxes #1.\y@vr}%
      \hrule height\y@linethick}\vskip-\y@linethick
  \fi}
\def\young(#1){%
  \y@setdim%
  \hskip\y@interspace%
  \y@lastargtest#1,
  \ifmmode\ify@vcenter\vcenter\fi\fi{%
  \vbox{\offinterlineskip
    \ify@islastarg\y@nelineabcboxes{#1}%
    \else\y@ungabc(#1)%
    \fi}}\hskip\y@interspace}
\def\y@ungabc(#1,#2){%
  \y@nelineabcboxes{#1}%
  \y@lastargtest#2,
  \ify@islastarg\y@nelineabcboxes{#2}%
  \else\y@ungabc(#2)%
  \fi}
\catcode`\@12\relax
 


\def\a{\alpha}
\cal{}

\def\half{{1\over 2}}

\font\cmss=cmss10
\font\cmsss=cmss10 at 7pt
\font\small=cmss10 at 3pt
\def\IL{\relax{\rm I\kern-.18em L}}
\def\IH{\relax{\rm I\kern-.18em H}}
\def\IR{\relax{\rm I\kern-.18em R}}
\def\inbar{\vrule height1.5ex width.4pt depth0pt}
\def\IC{\relax\hbox{$\inbar\kern-.3em{\rm C}$}}
\def\rlx{\relax\leavevmode}

\def\ZZ{\rlx\leavevmode\ifmmode\mathchoice{\hbox{\cmss Z\kern-.4em Z}}
 {\hbox{\cmss Z\kern-.4em Z}}{\lower.9pt\hbox{\cmsss Z\kern-.36em Z}}
 {\lower1.2pt\hbox{\cmsss Z\kern-.36em Z}}\else{\cmss Z\kern-.4em
 Z}\fi}
\def\IZ{\relax\ifmmode\mathchoice
{\hbox{\cmss Z\kern-.4em Z}}{\hbox{\cmss Z\kern-.4em Z}}
{\lower.9pt\hbox{\cmsss Z\kern-.4em Z}}
{\lower1.2pt\hbox{\cmsss Z\kern-.4em Z}}\else{\cmss Z\kern-.4em
Z}\fi}


\font\manual=manfnt
\def\dbend{\lower3.5pt\hbox{\manual\char127}}

\def\IZ{\relax\ifmmode\mathchoice
{\hbox{\cmss Z\kern-.4em Z}}{\hbox{\cmss Z\kern-.4em Z}}
{\lower.9pt\hbox{\cmsss Z\kern-.4em Z}} {\lower1.2pt\hbox{\cmsss
Z\kern-.4em Z}}\else{\cmss Z\kern-.4em Z}\fi}
\def\half {{1\over 2}}


\def\bar{\overline}

\def\rt2{\sqrt{2}}
\def\irt2{{1\over\sqrt{2}}}

\def\half{ {1\over 2}}


\Title{\vbox{\hbox{DISTA-2011}}} 
{\vbox{ 
\centerline{On Supergravity Amplitudes from Pure Spinor Strings}}}

\medskip
\centerline{P.~A. Grassi$^{~a,b,}$\foot{pgrassi@cern.ch} and
L. Sommovigo$^{~a,}$\foot{lsommovi@mfn.unipmn.it} }
\medskip 
\centerline{${a)}$ {\it DISTA, Universit\`a del Piemonte Orientale,} }
\centerline{\it via T. Michel, 11, 15121 Alessandria, ITALY.}
\medskip
\centerline{${b)}$ {\it INFN, Gruppo Collegato di Alessandria, Sezione di Torino, ITALY}}
\medskip
\vskip  .5cm
We use Pure Spinor String Theory to construct suitable building blocks to arrange the structure of supergravity amplitudes
with $n$-points and at higher loops in a very convenient way. Following Mafra {\it et al.}  proposal for $N=4$ super Yang-Mills amplitudes, we discuss its generalization to supergravity amplitudes. 
In particular, we first derive the building blocks for closed string amplitudes from closed vertices showing that, because of the factorization, several results of open string theory can be used; nevertheless, also new structures emerge. We provide a first list of these new ingredients and their BRST variations. Second, we analyze 4-point amplitudes showing how the different building blocks enter and finally we discuss the factorization of 5-point amplitudes in supergravity in terms of those structures. 

\Date{\ July 2011}

\lref\MafraJQ{
  C.~R.~Mafra, O.~Schlotterer, S.~Stieberger and D.~Tsimpis,
  arXiv:1012.3981 [hep-th].
}
\lref\MafraGJ{
  C.~R.~Mafra, O.~Schlotterer, S.~Stieberger, D.~Tsimpis,
Nucl.\ Phys.\  {\bf B846}, 359-393 (2011).
[arXiv:1011.0994 [hep-th]].
}

\lref\MafraWQ{
  C.~R.~Mafra,
  arXiv:0902.1552 [hep-th].
}

\lref\BernIA{
  Z.~Bern and T.~Dennen,
  arXiv:1103.0312 [hep-th].
}

\lref\MafraIR{
  C.~R.~Mafra,
  JHEP {\bf 1011}, 096 (2010)
  [arXiv:1007.3639 [hep-th]].
}

\lref\CarrascoMN{
  J.~J.~M.~Carrasco, H.~Johansson,
[arXiv:1106.4711 [hep-th]].
}
\lref\BernFY{
  Z.~Bern, J.~J.~M.~Carrasco, H.~Johansson,
Nucl.\ Phys.\ Proc.\ Suppl.\  {\bf 205-206}, 54-60 (2010).
[arXiv:1007.4297 [hep-th]].
}
\lref\MafraPN{
  C.~R.~Mafra,
[arXiv:1007.4999 [hep-th]].
}
\lref\MafraKJ{
  C.~R.~Mafra, O.~Schlotterer, S.~Stieberger,
[arXiv:1104.5224 [hep-th]].
}
\lref\MafraNV{
  C.~R.~Mafra, O.~Schlotterer, S.~Stieberger,
[arXiv:1106.2645 [hep-th]].
}
\lref\MafraNW{
  C.~R.~Mafra, O.~Schlotterer, S.~Stieberger,
[arXiv:1106.2646 [hep-th]].
}

\lref\BernYG{
  Z.~Bern, T.~Dennen, Y.~-t.~Huang, M.~Kiermaier,
Phys.\ Rev.\  {\bf D82}, 065003 (2010).
[arXiv:1004.0693 [hep-th]].
}
\lref\BernQJ{
  Z.~Bern, J.~J.~M.~Carrasco, H.~Johansson,
Phys.\ Rev.\  {\bf D78}, 085011 (2008). 
[arXiv:0805. 3993 [hep-ph]].
}
\lref\BjerrumBohrRD{
  N.~E.~J.~Bjerrum-Bohr, P.~H.~Damgaard, P.~Vanhove,
PRLTA,103,161602.\ 2009 {\bf 103}, 161602 (2009).
[arXiv:0907.1425 [hep-th]].
}
\lref\KleissNE{
  R.~Kleiss, H.~Kuijf,
Nucl.\ Phys.\  {\bf B312}, 616 (1989).
}
\lref\BianchiPU{
  M.~Bianchi, H.~Elvang, D.~Z.~Freedman,
JHEP {\bf 0809}, 063 (2008).
[arXiv:0805.0757 [hep-th]].
}


\newsec{Introduction}

Recently, a new method to compute super-Yang-Mills (SYM) amplitudes as a limit of string theory amplitudes within the Pure Spinor String Theory has been proposed by Carlos Mafra 
\MafraIR. The building blocks obtained from the OPE among the pure spinor string vertices are 
taken in the particle approximation (zero mode expansion). 
Then, using the zero mode saturation rules for the 
amplitudes one can select the non-vanishing ones. Moreover, the same rules also imply 
that the BRST-exact combinations drop out of the amplitudes. So, performing a study of the cohomology of those building blocks, a novel decomposition of the SYM amplitudes is proposed. It matches the tree level computations in \BernIA, pointing 
out some similarities and interesting relations among the amplitudes \refs{\BernQJ,\BjerrumBohrRD}. 

Afterwards, Mafra {\it et al.}   
in \refs{\MafraGJ,\MafraJQ} established a recursive formula simplifying the comparison with 
higher point amplitudes computed in the traditional way. This reorganization of diagrams, 
inspired by string theory, is very elegant and efficient, and it might shed some light on the relation between the perturbative expansion 
of SYM and SUGRA.\foot{The factorization at three level is rather well-established, while at higher orders it is still unproven \refs{\BernFY,\CarrascoMN}.} 
Further developments can be read in 
\refs{\MafraPN,\MafraKJ} and a complete analysis of string theory amplitudes and their 
reduction to field theory in pure spinor language is contained in papers \refs{\MafraNV,\MafraNW}.

Until now, only SYM amplitudes have been considered, but also 
the supergravity amplitudes require a parallel analysis in terms of corresponding building blocks  which may 
render the factorization of amplitudes in terms of open string building blocks apparent. 
In the present work, we apply the same technique to the  supergravity limit of superstrings.  
We provide the ingredients for the analysis along the lines of \MafraIR\ and we see that 
there are relevant differences. We analyze the amplitudes using these new building blocks. 

Our goal is to study the factorization of the supergravity amplitudes into 
SYM amplitudes
\refs{\BernIA,\BernYG}. In that respect, we refer to \BernYG\ where a complete and clear discussion on the factorization 
at tree level and at higher loops is provided. In particular, the view of gravity theory as the "square" of gauge theories is emphasized. 
For that,  we start from the factorized form of the vertices and we construct the building blocks, their BRST variations, the amplitudes and some relations. 
In addition, we derive some general theorems which do not take into account 
the factorized form of the vertices and we compute  the 4-  and 5-point functions. As pointed out in \refs{\MafraGJ,\MafraJQ} in order to evaluate higher point amplitudes 
new building blocks are needed. For 4-point functions one needs only two ingredients, but then by increasing the number of topologies and the number of external legs, one needs new emergent structures. We first discuss the building blocks for 4-point functions, their BRST variations (a complete cohomological analysis 
should follow from work \refs{\MafraNV,\MafraNW}) and their relations. We discuss some relevant issues such as the labeling of closed building blocks in terms 
of the factorized form. Then, moving towards 5-point functions we show that new structures are indeed needed and their BRST variations call for these new objects in order to guarantee the closure of the BRST algebra. 

In sec. 2, we briefly review Mafra's analysis of open string building blocks. In sec. 3, we derive closed string building blocks for supergravity amplitudes, we analyze their BRST transformation rules, the factorization and those building blocks that do not factorize. In sec. 4, the pole structure, the form of the amplitudes and the kinematic factor are discussed. In sec. 5 we review SYM amplitudes and, finally, in sec. 6 we compute supergravity amplitudes and we discuss the factorization properties. In app. A the BCJ relations for closed amplitude are discussed. 

\newsec{Open String Building Blocks}

In this section we give a brief description of the open string building blocks. 
According to Mafra \MafraJQ\ we need two different types of superfields: the bosonic ones, $E$, with ghost number 2, and the fermionic ones, ${\cal M}$, with ghost number 1. 

They are derived from the unintegrated vertex operator $V_i$ with the integrated one $U_i$ (the index $i$ labels the background fields appearing in the vertex operator) by considering their short distance expansion
\eqn\primaA{
V_i(z) U_j(w) \rightarrow {1\over (z-w)} T_{ij}\,.
}
The new superfield $T_{ij}$ is a fermionic ghost number 1 vertex which depends upon the 
gluon polarization, the gluino wave function and upon the momenta $k_i$ and $k_j$ of the lines 
$i$ and $j$ of an amplitude. Acting on both sides of \primaA\ with the BRST charge, we get 
\eqn\primaB{
Q T_{ij} = - s_{ij} V_i V_j\,,
}
where $s_{ij} = k_i \cdot k_j$ is the first Maldestam invariant. Defining ${\cal M}_i \equiv V_i$, 
we can construct the $E$'s and the ${\cal M}$'s using the following relations:
\eqn\blockA{E_{i_1 \dots i_n} = \sum_{p=1}^{n-1} {\cal M}_{i_1 \dots i_p} \, {\cal M}_{i_{p+1} \dots i_n}\,,} 
\eqn\blockB{QE_{i_1 \dots i_p} = 0 , \quad E_{i_1 \dots i_p} = Q {\cal M}_{i_1 \dots i_p} \; {\rm if} \; s_{i_1 \dots i_p} \neq 0\,,}
where $s_{i_1 \dots i_p} = \half (k_1 + \dots + k_p)^2$ are the Mandelstam invariants. Let us give a few explicit examples; by taking $p=2$ we have:
$$
E_{ij} = {\cal M}_{i} {\cal M}_j = Q {\cal M}_{ij}\,.
$$
${\cal M}_{ij}$ is clearly related to the superfields $T_{ij}$ by ${\cal M}_{ij} = - {{T_{ij}}\over{s_{ij}}}$ (if $s_{ij} \neq 0$). The case $p=3$ is governed by the superfield $T_{ijk}$, in terms of which we write the ${\cal M}_{ijk}$; first of all, one defines
\eqn\blockC{E_{ijk} = {\cal M}_i {\cal M}_{jk} + {\cal M}_{ij}{\cal M}_k = Q {\cal M}_{ijk}\,,}
then, since the superfield $T_{ijk}$ satisfies the following relations \refs{\MafraNV,\MafraNW}:
\eqn\openfiveB{
Q T_{ijk} = s_{ij} T_{jk} V_{i} + s_{ij} T_{ki} V_{j} - (s_{ki}+s_{jk}) T_{ij} V_{k} \,,
}
$$
T_{ijk} + T_{jki} + T_{kij} = Q \chi_{ijk} \,,
$$
a short computation shows that 
\eqn\blockD{
{\cal M}_{ijk} = {1 \over s_{ijk}} \left( {T_{ijk} \over s_{ij} } - {T_{jki} \over s_{jk}} \right)\,,
}
has the desired properties if $s_{ijk} \neq 0$. Proceeding this way, it is possible to reconstruct 
the tower of ${\cal M}$'s and and ${ E}$ 
needed in order to build $n$-point amplitudes for open strings.

\newsec{Closed String Building Blocks}

We repeat here a similar construction for closed string building blocks, pointing out the differences with the open string case and their common features. 

The vertex operators in closed string theory are classified in terms 
of the ghost number and the conformal weight. We work in a factorable background for 
which we can keep the (anti)-holomorphicity manifest at every stage. This guarantees the factorization of the vertex operators if the polarizations of the background fields are factorized. 
So, we have the following set of vertices 
\eqn\ClosedAA{
V^{(0|0)}_{z\bar z}\,, \quad
U^{(0|1)}_{z}\,, \quad  
U^{(1|0)}_{\bar z}\,, \quad  
\Omega^{(1|1)}\,,
}
which satisfy the descent equations (the indices $z, \bar z$ are left explicit in order to remind the reader the conformal weight of each single vertex operators)
\eqn\ClosedAB{
[Q_L, V^{(0|0)}_{z\bar z}] = \partial_{z}  U^{(1|0)}_{\bar z}\,, \quad
[Q_R, V^{(0|0)}_{z\bar z}] = \partial_{\bar z}  U^{(0|1)}_{z}\,, \quad
}
$$
\{Q_L, U^{(0|1)}_{z}\} = \partial_z \Omega^{(1|1)}\,, \quad
\{Q_R, U^{(1|0)}_{\bar z}\} = \partial_{\bar z} \Omega^{(1|1)}\,, \quad
$$
$$
[Q_L,  \Omega^{(1|1)}] = 0\,, \quad
[Q_R,  \Omega^{(1|1)}] = 0\,,
$$
where $Q_L$ and $Q_R$ are the left- and right-moving BRST charges, respectively 
\lref\BerkovitsUE{
  N.~Berkovits and P.~S.~Howe,
  Nucl.\ Phys.\  B {\bf 635}, 75 (2002)
  [arXiv:hep-th/0112160].
}
\lref\GrassiIH{
  P.~A.~Grassi, L.~Tamassia,
JHEP {\bf 0407}, 071 (2004).
[hep-th/0405072].
}
\refs{\BerkovitsUE,\GrassiIH}. 
They satisfy $Q^2_L=Q^2_R = \{Q_L, Q_R\}=0$.  
We can see the vertices \ClosedAA\ as formed by taking 
the tensor products of the left- and right-moving vertex operators 
\eqn\ClosedAC{
V^{(0|0)}_{z\bar z} = V^{L, (0)}_z \otimes V^{R, (0)}_{\bar z}\,, \quad
U^{(0|1)}_{z} = V^{L, (0)}_z \otimes U^{R,(1)} \,, \quad  
}
$$
U^{(1|0)}_{\bar z} = U^{L, (1)}\otimes V^{R, (0)}_{\bar z} \,, \quad  
\Omega^{(1|1)} = U^{L, (1)} \otimes U^{R, (1)}\,,
$$
where $V^{L, (0)}_z, V^{R, (0)}_{\bar z}$ are the integrands of the integrated vertices and 
$U^{L, (1)}, U^{R, (1)}$ are the unintegrated vertices for the left- and right-mover sector, respectively. 
Now, we use the results of \MafraWQ\ to construct new building blocks for closed strings, in particular we fuse the unintegrated and the integrated vertex operator 
$$
U^{L,(1)}(z) V^{L,(0)}_w(w) \rightarrow {L^{L,(1)}_{z} \over (z-w)}\,, \quad\quad
U^{R,(1)}(\bar z) V^{R,(0)}_{\bar w}(\bar w) \rightarrow {L^{R,(1)}_{\bar z} \over (\bar z-\bar w)}\,, 
$$
into new objects $L^{L,(1)}_{z}$ and $L^{R,(1)}_{\bar z}$. Their BRST variations can be easily deduced 
by assuming the form $U^{L,(1)}(z) = U^{L,(1)}(\theta, \bar \theta) e^{i k\cdot X}$ where 
$U^{L,(1)}(\theta, \bar \theta)$ is a superfield of $\theta, \bar\theta$ and depends upon the pure spinor 
variables. 

Then, by performing the same operation on the closed string vertex operators 
we have the useful new blocks
\eqn\ClosedAD{
U^{(0|1)}_{z}(z, \bar z)  U^{(1|0)}_{\bar w}(w, \bar w) \rightarrow 
{{\cal K}^{(1|1)} \over |z-w|^2}\,,
} 
$$
U^{(0|1)}_{z}(z, \bar z) \Omega^{(1|1)}(w, \bar w) \rightarrow {{\cal J}^{(1|2)} \over (z-w)}\,, \quad
U^{(1|0)}_{\bar z}(z, \bar z) \Omega^{(1|1)}(w, \bar w) \rightarrow {\overline{\cal J}^{(2|1)} \over (\bar z-\bar w)}\,.
$$

Now, we have to discuss the index assignment. 
Merging left- and right-movers we have to clarify one point: given the vertex operator 
for left-movers $U^{L,(1)}_i(\epsilon^i_m, u^i_\alpha, p^i_m)$, where 
$\epsilon^i_m, u^i_\alpha, p^i_m$ are the gluon polarization, the gluino spinorial wave function and the momentum, we set $\Omega^{(1|1)}_{iu} = U^{L, (1)}_i 
\otimes U^{R, (1)}_u$.  
Therefore, this closed string vertex operator depends upon the background fields 
$$
g^{iu}_{mn} + \eta_{mn} \phi = \epsilon^i_{(m} \epsilon^u_{n)}\,, \quad
b^{iu}_{mn} = \epsilon^i_{[m} \epsilon^u_{n]}\,,
$$
$$
\psi^{iu}_{\alpha n} =  \epsilon^u_{n} u^i_{\alpha}\,, \quad
 \psi^{iu}_{m\hat\alpha} =  \epsilon^i_{m}  u^u_{\hat  \alpha}\,, \quad
f^{iu}_{\alpha\hat \beta} = u^i_{\alpha} u^u_{\hat \beta},
$$
where $f^{iu}_{\alpha\hat \beta}$ are the field strengths for the RR fields. The momenta for the two merging vertices $p_m^i$ and $p_n^u$ must be the same (unless winding modes are considered) and therefore Mafra's building blocks ${\cal K}^{(1|1)}, {\cal J}^{(1|2)}$ and 
$\overline{\cal J}^{(2|1)}$ acquire two pairs of indices which we denote as 
${\cal K}^{(1|1)} \rightarrow {\cal K}^{(1|1)}_{ij|uv}$, or, in a more compact way, as 
${\cal K}_{IJ}^{(1|1)}$ where $I$ denote the pair $(iu)$ and $J$ the pair $(jv)$. In the following, 
we exhibit some expressions which do not rely on the factorization of the vertices and which may be writable only in terms of the indices $I,J,K,....$. 

A remark: if we define the closed string vertex operator $\Omega^{(1|1)}_{iu} = U^{L, (1)}_i \otimes U^{R, (1)}_u$ from tensoring left- and right-movers, we have the following identities 
$$
\Omega^{(1|1)}_{iu} \Omega^{(1|1)}_{jv} = U^{L, (1)}_i \otimes U^{R, (1)}_u \,
U^{L, (1)}_j \otimes U^{R, (1)}_v =
$$
$$
= - U^{L, (1)}_i \otimes U^{R, (1)}_v \,
U^{L, (1)}_j \otimes U^{R, (1)}_u  = - \Omega^{(1|1)}_{iv} \Omega^{(1|1)}_{ju} \,.
$$
Some other relations among the building blocks of Mafra {\it et al.}'s construction  follow 
from this idea. 

\subsec{BRST Variations}

We derive the BRST transformations of the closed string theory building blocks. 
First we present the derivation based on the factorized form of the vertices (and this allows us to determine some useful relations among the building blocks) and then we provide the BRST variations in terms of the compact expressions. 

\subsec{Factorized Building Blocks}

For closed string we use the following building blocks
\eqn\CloseA{
\Omega^{(1|1)}_{iu} \,, \quad
{\cal K}^{(1|1)}_{ij|uv} \,, \quad
{\cal J}^{(1|2)}_{ij|uv} \,, \quad 
\overline{\cal J}^{(2|1)}_{ij|uv} \,, \quad 
}
which are written in terms of factorized expressions 
\eqn\ClosedD{
\Omega^{(1|1)}_{iu} = V^L_i \otimes V^R_{u}\,, \quad\quad
{\cal J}^{(1|2)}_{ij|uv}  = {\cal M }^L_{ij} \otimes V^R_{u} V^R_{v}\,, 
}
$$
\overline{\cal J}^{(2|1)}_{ij|uv}  = V^L_{i} V^L_{j} \otimes {\cal M}^R_{uv} \,, \quad\quad
{\cal K}^{(1|1)}_{ij|uv} = {\cal M}^L_{ij} \otimes {\cal M}^R_{uv}\,. 
$$

The variations under BRST $Q_L$ and $Q_R$ of \CloseA\ are
\eqn\CloseA{
Q_L\Omega^{(1|1)}_{iu} =0 \,, \quad
Q_L{\cal K}^{(1|1)}_{ij|uv} = \overline{\cal J}^{(2|1)}_{ij|uv}  \,, \quad
Q_L{\cal J}^{(1|2)}_{ij|uv} = - \Omega^{(1|1)}_{iu}  \Omega^{(1|1)}_{jv}  \,, \quad 
Q_L\overline{\cal J}^{(2|1)}_{ij|uv} =0\,, \quad 
}
and 
\eqn\CloseB{
Q_R\Omega^{(1|1)}_{iu} =0 \,, \quad
Q_R{\cal K}^{(1|1)}_{ij|uv} = - {\cal J}^{(1|2)}_{ij|uv}  \,, \quad
Q_R{\cal J}^{(1|2)}_{ij|uv} =0  \,, \quad 
Q_R\overline{\cal J}^{(2|1)}_{ij|uv} = - \Omega^{(1|1)}_{iu}  \Omega^{(1|1)}_{jv}\,.
}
Moreover, the following identities hold 
$$
\Omega^{(1|1)}_{iu}  \Omega^{(1|1)}_{jv} = - 
\Omega^{(1|1)}_{iv}  \Omega^{(1|1)}_{ju} = - 
\Omega^{(1|1)}_{ju}  \Omega^{(1|1)}_{iv} = 
\Omega^{(1|1)}_{jv}  \Omega^{(1|1)}_{iu}\,, 
 $$
 $$
{\cal J}^{(1|2)}_{ij|uv}  = -  {\cal J}^{(1|2)}_{ij|vu} = - {\cal J}^{(1|2)}_{ji|uv} + Q_L {\cal X}_{ij|uv}\,,
$$
$$
\overline{\cal J}^{(2|1)}_{ij|uv}  = -  \overline {\cal J}^{(2|1)}_{ji|vu} = - 
\overline{\cal J}^{(2|1)}_{ij|vu} + Q_R \overline{\cal X}_{ij|uv}\,. 
 $$
These relations are useful for constructing the amplitudes. Furthermore, to compute
the BRST variations of the amplitudes we have to take into account also 
the following relations
$$
\Omega^{(1|1)}_{iu} {\cal J}^{(1|2)}_{ij|wv} = - \Omega^{(1|1)}_{iw} {\cal J}^{(1|2)}_{ij|uv}\,, \quad
\Omega^{(1|1)}_{iu} {\cal J}^{(1|2)}_{ij|wv} = - \Omega^{(1|1)}_{iv} {\cal J}^{(1|2)}_{ij|uw}\,,
$$
which can be easily deduced given the definitions \ClosedD.

\subsec{Non-factorized Building Blocks}

Now, we give the same expressions in a more compact way. 
We have the following building blocks 
\eqn\bbA{
{\cal K}^{(1|1)}_{IJ}\,, \quad
{\cal J}^{(1|2)}_{IJ}\,, \quad
\bar{\cal J}^{(2|1)}_{IJ}\,, \quad
{\Omega}^{(1|1)}_{I}\,, \quad
}
which transform as follows
\eqn\bbB{
Q_L {\cal K}^{(1|1)}_{IJ} = \bar{\cal J}^{(2|1}_{IJ}\,, \quad\quad 
Q_L \bar{\cal J}^{(2|1)}_{IJ}=0\,,
}
$$
Q_R {\cal K}^{(1|1)}_{IJ} = - {\cal J}^{(1|2}_{IJ}\,, \quad\quad
Q_R {\cal J}^{(1|2)}_{IJ} =0\,, 
$$
$$
Q_R  \bar{\cal J}^{(2|1)}_{IJ} = \Omega^{(1|1)}_I \Omega^{(1|1)}_J\,, 
Q_L  {\cal J}^{(2|1)}_{IJ} = \Omega^{(1|1)}_I \Omega^{(1|1)}_J\,.
$$

For non-factorable building blocks, in order to prove the BRST invariance, we have to 
do something more since we need to explore some relations among the building blocks, which 
appear to be trivial in the factorized form but not  in the non-factorized form. By considering the short distance expansion of 
$\Omega^{(1|1)}_I(z) \Omega^{(1|1)}_J(w) U_k^{(1|0)}(x)$ 
we get the following equation 
\eqn\bbC{
(\bar w - \bar x) \Omega_J(w) \bar{\cal J}^{(2|1)}_{IK}(x) = 
(\bar z - \bar x) \Omega_I(z) \bar{\cal J}^{(2|1)}_{JK}(x) \,, 
}
from which we deduce that 
$$\Omega_J \bar{\cal J}^{(2|1)}_{IK} = 
\Omega_I \bar{\cal J}^{(2|1)}_{JK}\,,$$  
which turns out to be fundamental in the following. Also the complex conjugate 
equation is valid. 

\subsec{Non-factorized Higher-Point Bulding Blocks}
These equations are sufficient for the 4-point functions, but for the 5-point functions one needs 
additional building blocks. Here we give them with their variations
\eqn\CCa{
Q_L {\cal K}^{(1|1)}_{IJK} = ({\cal M}^L_{ij} {\cal M}^L_{k} + {\cal M}^L_{i} {\cal M}^L_{jk}) 
\otimes {\cal M}^R_{uvw} = {\cal J}^{(2|1)}_{I [JK]} + 
{\cal J}^{(2|1)}_{[I J]K} \,,
}
where ${\cal K}^{(1|1)}_{IJK} $ is defined as follows
\eqn\KKtre{
{\cal K}^{(1|1)}_{IJK} \equiv {\cal M}_{ijk} \otimes {\cal M}_{uvw}\,,
}
and
$$
Q_R {\cal K}^{(1|1)}_{IJK} = - {\cal Y}^{(1|2)}_{I [JK]} - {\cal Y}^{(1|2)}_{[I J]K} \,,
$$
$$
Q_L {\cal J}^{(2|1)}_{I[JK]} = -  {\cal M}^L_{i} {\cal M}^L_{j}{\cal M}^L_{k} \otimes 
{\cal M}^R_{uvw}  = - \Theta^{(3|1)}_{IJ K}\,, 
$$
$$
Q_L {\cal J}^{(2|1)}_{[IJ]K]} =  \Theta^{(3|1)}_{IJK}\,,
$$
$$
Q_R {\cal Y}^{(1|2)}_{[IJ]K} = - \Xi^{(1|3)}_{IJK}\,, 
$$
$$
Q_R {\cal Y}^{(1|2)}_{I[JK]} =  \Xi^{(1|3)}_{IJK}\,,  
$$
$$
Q_R {\cal J}^{(2|1)}_{I[JK]} =  {\cal M}^L_{ij} {\cal M}^L_{k} \otimes 
({\cal M}^R_{uv} {\cal M}^R_{w} + {\cal M}^R_{u} {\cal M}^R_{vw})  = 
-\Omega^{(1|1)}_I {\cal K}^{(1|1)}_{IJ} + K^{(1|1)}_{IJ} \Omega^{(1|1)}_K + Q_L {\cal Y}^{(1|2)}_{I[J K]}\,, 
$$
$$
Q_R {\cal J}^{(2|1)}_{[IJ]K} =  {\cal M}^L_i {\cal M}^L_{jk} \otimes 
({\cal M}^R_{uv} {\cal M}^R_{w} + {\cal M}^R_{u} {\cal M}^R_{vw})  = 
\Omega^{(1|1)}_I {\cal K}^{(1|1)}_{IJ} - K^{(1|1)}_{IJ} 
\Omega^{(1|1)}_K + Q_L {\cal Y}^{(1|2)}_{[IJ] K}\,, 
$$
$$
Q_L \Theta^{(3|1)}_{IJK} =0\,, 
$$
$$
Q_R \Theta^{(3|1)}_{IJK} = - \Omega^{(1|1)}_{I} \bar{\cal J}^{(2|1)}_{JK} - \bar{\cal J}^{(2|1)}_{IJ}  \Omega^{(1|1)}_{K}\,,
$$
$$
Q_L \Xi^{(1|3)}_{IJK} = \Omega^{(1|1)}_{I} {\cal J}^{(1|2)}_{JK} + {\cal J}^{(1|2)}_{IJ}  \Omega^{(1|1)}_{K}\,,
$$
$$ 
Q_R \Xi^{(1|3)}_{IJK} = 0\,.
$$
It can be easily verified that $Q_L^2 = Q_R^2= \{Q_L, Q_R\}=0$. 
In order to disentangle $ Q_R {\cal J}^{(2|1)}_{I[JK]}$ from $Q_L {\cal Y}^{(1|2)}_{I[J K]}$ we introduce the new blocks 
\eqn\CCb{
Q_L {\cal Y}^{(1|2)}_{I[J K]} = \Delta^{(2|2)}_{I[JK]}\,, \quad \quad 
Q_L {\cal Y}^{(1|2)}_{[IJ] K} = \Delta^{(2|2)}_{[IJ]K}\,, 
}
which, by consistency, transform as follows
\eqn\CCc{
Q_R \Delta^{(2|2)}_{[IJ] K} =\Omega^{(1|1)}_{I} {\cal J}^{(1|2)}_{JK}  - {\cal J}^{(1|2)}_{IJ}  \Omega^{(1|1)}_{K}\,,
\quad\quad
Q_R \Delta^{(2|2)}_{I[J K]} =  -\Omega^{(1|1)}_{I} {\cal J}^{(1|2)}_{JK} + {\cal J}^{(1|2)}_{IJ}  \Omega^{(1|1)}_{K}\,.
}

In the following, we show how these building blocks are used in the construction of 5-point functions. As in SYM case, adding a new external leg implies new structures to be considered. 

\newsec{Pole Structure, Amplitudes and Kinetic Factors}

Here, we discuss different types of structures appearing in the amplitudes. 
In particular, we consider the following structure for the 
$n$-point amplitude
\eqn\struA{
{\cal A}^{open}_{n} = \sum_{\a} { c_\a n_\a \over \prod_{i_\a} s_{i_\a}}\,.
}
The sum is extended over all possible diagrams at a given loop order for the amplitude 
${\cal A}^{open}_{n}$; $c_\a$ is the color factor, 
$n_\a$ is the kinematical factor and $\prod_{i_\a} s_{i_\a}$ is the 
product of the propagators $s_{i_\a}$ of the diagram $\a$. The color factors are extracted by selecting the 
color matrices of each vertex and therefore we can denote by ${\cal A}^{open}_{n}(1,\dots,n)$ 
the amplitude after the extraction of its color factor. Mafra {\it et al.} proposed 
to identify the $n_\alpha$ with the building blocks
\eqn\struB{
n_\a = \langle {\cal N}_\alpha \rangle = \langle V_{i_1} \dots V_{i_n} T_{j_1 j_2} \dots  T_{j_m j_m} \dots \rangle\,.  
}
If $n_\alpha$ represents the physical amplitudes they must be well defined in terms 
of BRST invariant building blocks and this can be checked applying directly the variations 
of expression  ${\cal N}_\alpha $ inside of the angular brackets in \struB. 

The variation of the building blocks $T_{i_1 \dots i_n}$ is proportional to 
$s_{i_1 \dots i_n}$ and they cancel with the poles in \struA. Notice that, for a given loop order, the number of 
propagators $s_{i_\a}$ is fixed for each diagram and therefore the possible types of building blocks are 
also fixed. Therefore, only combinations of \struB\ with rational numbers are needed in order to 
obtain BRST invariant quantities.  If we insist on having a BRST combination of $n_\a$, independently 
of the pole structure, we have, for each independent invariant $s_{ij}$, an independent equation. 

A general form for  ${\cal N}_\alpha $, for an $n$-point function at tree level, can be established 
\eqn\struC{
{\cal N}_\a= 
\a^{i_1 \dots i_n} V_{i_1} \dots V_{i_p} T_{i_{p+1} i_{p+2}} \dots T_{i_{q_1} i_q} \dots T_{i_{n-r} \dots i_n}\,.
}
Since the total ghost number must be 3 for tree level and since all building blocks have ghost number 
$+1$ we have at most three structures. They also have to match the number of points of the amplitude, 
namely the number of indices $i_1, \dots, i_n$ must be $n$, and they should be suitably distributed into the different building blocks. 
For example, for  a 6-point functions, we could have $V_i V_j T_{klmn}$, or 
$V_i T_{jk} T_{lmn}$. For 7-point functions we could have $V_i V_j T_{klmno}$, 
$V_i T_{jl} T_{kmno}$, $V_i T_{jkl} T_{mno}$ and so on. In principle, we could not exclude any of them.\foot{In \MafraKJ,  
a general decomposition is proposed for the amplitudes $n_\a$.} 
Notice that we have to remove those terms which are BRST exact:  all possible expressions with ghost number 2 
which could cancel the BRST closed amplitudes. However, 
since we are looking at expressions with ghost number 3, the BRST exact have ghost 2, but in addition we have ghost-for-ghost expressions which remove some of the BRST exact terms and they have ghost number 1. So, we are looking for cohomologies of the following types
\eqn\struD{
Q  {\cal N}_\alpha^{(3)}=0\,,\quad\quad
\delta  {\cal N}_\alpha^{(3)} = Q  {\cal N}_\alpha^{(2)}\,, \quad\quad
\delta  {\cal N}_\alpha^{(2)} = Q  {\cal N}_\alpha^{(1)}\,. 
}
Notice that we do not have any ghost-number zero quantity and therefore the ladder of quantities stops 
at ${\cal N}_\alpha^{(1)}$. In addition, for any $n$-point amplitude always exists a structure 
$T^{(1)}_{i_1 \dots i_n}$. 

For constructing the closed amplitudes we can replace $c_\a$ and $n_\a$ with 
the two integrand for the left- and right-movers, which we are going to denote 
${\cal N}_{\a_L}$ and ${\cal N}_{\a_R}$ and the expression 
for the amplitude becomes
\eqn\struA{
{\cal A}^{closed}_{n} = \sum_{\a}  \Big\langle 
{ ({\cal N}^{(3,0)}_{\a_L} {\cal N}^{(0,3)}_{\a_R} + {\cal N}^{(3,0)}_{\a_L} {\cal N}^{(0,3)}_{\a_R}) \over \prod_{i_\a} s_{i_\a}} \Big\rangle\,.
}
Again, we can apply the above considerations to derive the BRST invariant amplitudes. Here 
we have to notice the main difference with the open string amplitude: the cohomologies 
involved in computation are of the following type: 
\eqn\struE{
Q_R\Big( { {\cal N}^{(0,3)}_{\a_R} \over  \prod_{i_\a} s_{i_\a}} \Big)=0\,, \quad
\delta {\cal N}^{(0,3)}_{\a_R} = Q_R \Big({\cal N}^{(0,2)}_{\a_R} { F(s_{i_\a}) \over  \prod_{i_\a} s_{i_\a}} \Big)\,,
\quad
\delta {\cal N}^{(0,2)}_{\a_R} = Q_R \Big({\cal N}^{(0,1)}_{\a_R} { G(s_{i_\a}) \over  \prod_{i_\a} s_{i_\a}} \Big)\,,
}
where $F(s_{i_\a})$ and $G(s_{i_\a})$ are linear and quadratic expressions in momenta 
$s_{i_\a}$. Since the variations of the expression ${\cal N}^{(0,r)}_{\a_R}$ produce a zero in the external momenta, we need the poles to cancel the momenta in the numerators. Vice-versa, in the case of 
${\cal N}^{(r,0)}_{\a_L}$, we solve the cohomology 
\eqn\struE{
Q_L \Big( {\cal N}^{(3,0)}_{\a_L} \Big)=0\,, \quad\quad 
\delta {\cal N}^{(3,0)}_{\a_L} = Q_L \Big( {\cal N}^{(2,0)}_{\a_L} \Big)\,,
\quad\quad
\delta {\cal N}^{(2,0)}_{\a_L} = Q_L \Big({\cal N}^{(1,0)}_{\a_L} \Big)\,,
}
which is different from the equation above in the fact that the BRST variations of the various terms 
must cancel the poles in the denominator. 
In the reference-list papers, only the first type of cohomologies are computed. 

\newsec{Open Amplitudes}

For the 4-point functions\foot{In the following we use the terminology "amplitudes" to denote both the correlation functions and 
the integrand. We distinguish them by the notation ${\cal A}^{open/closed}_{n-point}$ for the amplitudes and ${\cal N}^{open/closed}_{n-point}$ for the integrand.}, 
we start from the ansatz
\eqn\openB{
{\cal N}^{open}_4 = \alpha^{[ij] [kl]} {V_i^{(1)} V_j^{(1)} T_{kl}^{(1)} \over s_{kl} }\,. 
}
Requiring BRST closure, we get the condition
\eqn\openBA{
\alpha^{[ij kl]} =0\,.
}  
The coefficients $\alpha^{[ij] [kl]}$ are antisymmetric in the exchange of the indices 
in the pairs  and therefore they can be decomposed into three pieces 
$$
\matrix{{\cmsss \yng(1,1)}} \quad \otimes \quad \matrix{\cmsss \yng(1,1)} \quad=\quad 
\matrix{\cmsss \yng(2,2)} \quad\oplus \quad \matrix{\cmsss \yng(2,1,1)}\quad \oplus \quad 
\matrix{\cmsss \yng(1,1,1,1)}\,.
$$
Actually we are not interested in all the tensors deriving from the above diagrams, because we only need those in which all the indices take different values (different legs are labeled by different numbers). Hence we have to evaluate the dimension of the diagrams with no repetitions, rather than that of the usual ones. This can be done by noticing that, taking this condition into account, the dimension of the totally symmetric diagram is the same of the totally antisymmetric one, the others following as usual. For instance, 
$$
{\rm{Dim} \matrix{\cmsss \yng(1,1,1,1)}} = {\rm{Dim} \matrix{\cmsss \yng(4)}} = {N (N-1) (N-2) (N-3) \over 4!}\,,
$$
$$
{\rm{Dim} \matrix{\cmsss \yng(3,1)}} = {\rm{Dim} \matrix{\cmsss \yng(2,1,1)}} = {N (N-1) (N-2) (N-3) \over 8}\,,
$$
$$
{\rm{Dim} \matrix{\cmsss \yng(2,2)}} = {N (N-1) (N-2) (N-3) \over 12}\,,
$$
which, in the case of four external legs, have dimensions 1, 1, 3, 3 and 2 respectively.  
Therefore, in the amplitude ${\cal N}^{open}_4$ only the former two diagrams are involved, the latter being removed by BRST symmetry. However, 
while in the first case (for the Young tableaux ${\small \yng(2,2)}$) the amplitude cannot be 
written as a BRST variation, for the second tableaux ${\small \yng(2,1,1)}$ we have 
\eqn\openC{
{\cal N}^{open}_4 = \alpha^{[ij] [kl]}_{\small \yng(2,1,1)} {(Q T^{(1)}_{ij})  T_{kl}^{(1)} \over s_{ij} s_{kl}} = {1 \over 2} Q \Big( 
\alpha^{[ij] [kl]}_{\small \yng(2,1,1)}  {T^{(1)}_{ij}  T_{kl}^{(1)} \over s_{ij} s_{kl} }\Big) \,,
}
which is non zero since the coefficients $\alpha^{[ij] [kl]}_{\small \yng(2,1,1)} $ are antisymmetric in the exchange 
of the pairs of indices, so that we can conclude that there are two independent amplitudes. Indeed, the expressions $T_{kl}^{(1)}$ are anticommuting and $s_{ij}= s_{kl}$.  

Let us move to 5-pt function. We know that the amplitude can be written in terms of the $T_{ij}$ blocks as:
\eqn\openfiveA{
{\cal N}^{open}_5 =
\alpha^{m [[ij] [kl]]} {V_m T_{ij} T_{kl} \over s_{ij} s_{kl} }\,, 
}
where the coefficient $\alpha$ is antisymmetric in the exchange of $ij$, $kl$ and of the couples $ij$ with $kl$. Therefore it belongs to the following representation
$$
\matrix{{\cmsss \yng(1,1)}} \quad \otimes \quad \matrix{\cmsss \yng(1,1)} \quad \otimes \quad \matrix{\cmsss \yng(1)} = \quad \left( \matrix{\cmsss \yng(2,2)} \quad \oplus \quad \matrix{\cmsss \yng(2,1,1)} \quad \oplus \quad \matrix{\cmsss \yng(1,1,1,1)} \right) \quad \otimes \quad \matrix{\cmsss \yng(1)}\,,
$$
where the first and the last terms in brackets do not contribute, because they are symmetric in the exchange of the pairs; we are then left with the following representations
$$
\matrix{\cmsss \yng(3,1,1)} \quad \oplus \quad \matrix{\cmsss \yng(2,2,1)} \quad \oplus \quad \matrix{\cmsss \yng(2,1,1,1)}\,,
$$
with dimension 6, 5 and 4 respectively. This amplitude is BRST invariant if $\alpha^{m [[ij] [kl]]} + \alpha^{i [[jm] [kl]]} + \alpha^{j [[mi][kl]]} = 0$, that is when $\alpha$ is antisymmetrized in the first 3 indices. Therefore, all the $\alpha$'s which are in the product of a 3-indices antisymmetric times a 2-indices antisymmetric are canceled by BRST invariance, 
$$
\matrix{\cmsss \yng(1,1,1)} \quad \otimes \quad \matrix{\cmsss \yng(1,1)} \quad = \quad \matrix{\cmsss \yng(1,1,1,1,1)} \quad \oplus \quad \matrix{\cmsss \yng(2,2,1)} \quad \oplus \quad \matrix{\cmsss \yng(2,1,1,1)}\,,
$$ 
so that the term corresponding to ${\small \yng(3,1,1)}$, which has dimension 6, vanishes.
Then, we can ask whether this amplitude is BRST exact. This is only possible if we use the additional building block $T^{(1)}_{ijk}$ whose BRST variation is given in \openfiveB. 
Indeed, we have
\eqn\openfiveC{
{\cal N}^{open}_5 = Q \Big(  \alpha^{[ij] kl m} {\cal M}^{(1)}_{ij} {\cal M}^{(1)}_{klm}\Big)\,. 
}

Notice that we can rewrite \openfiveA\ using $T_{ijk}$ (or better using ${\cal M}_{ijk}$) by the 
following identity
\eqn\openfiveCA{
\alpha^{[[ij][kl]] m} {\cal M}_{ij} {\cal M}_{kl} {\cal M}_m = {1\over 2} \alpha^{[[ij][kl]] m} 
 {\cal  M}_{ij}Q {\cal M}_{klm} =}
$$
= Q \Big({1\over 2} \alpha^{[[ij][kl]] m} {\cal M}_{ij} {\cal M}_{klm}\Big) +
{1\over 2} \alpha^{[[ij][kl]] m} {\cal M}_i {\cal M}_j {\cal M}_{klm}
$$
which is possible since we sum up over all possibilities. This means that we can cancel only some combinations of the diagrams  ${\small \yng(2,2,1)}$ and  ${\small \yng(2,1,1,1)}$
 and the remaining ones are indeed element of the cohomology. 
 
We have to clarify the relation between 
\openB\ and the expressions given in Mafra's paper. Indeed, he found that there are 3 non-trivial 
BRST invariant amplitudes for 4-pt function. They are labeled as follows 
${\cal A}(1,2,3,4)\,, {\cal A}(1,3,2,4)\,, {\cal A}(1,4,2,3)$. This means that we are making a specific choice of the coefficients $\alpha^{[ij] [kl]}$ which reads
\eqn\openD{
\alpha^{[ij] [kl]}(a,b,c,d) = \delta_a^{[i} \delta_b^{(j]} \delta_c^{[k)} \delta_d^{l]}\,. 
}
 With this ansatz, we find that 
\eqn\openE{
{\cal N}^{open}_4(1,2,3,4) = {\cal A}(1,2,3,4) - {\cal A}(1,3,2,4)\
}

As discussed above, we have to study another cohomology, namely the cohomology for kinetic terms to be used for closed amplitudes. this can be achieved by starting from a different ansatz of the form 
\eqn\kinoA{
{\cal N}^{open}_5 = \Big( {\alpha_1^{m [[ij] [kl]]} \over s_{ij} } + 
{\alpha_2^{m [[ij] [kl]]} \over s_{kl}} \Big) V_m T_{ij} T_{kl} 
\,, 
}
where we changed the pole structure such that the variation has no pole in $s$. It can be 
easily checked that it exists a solution for the coefficients $\alpha_1^{m [[ij] [kl]]}$ and 
$\alpha_2^{m [[ij] [kl]]}$ which is independent of $s_{ab}$.  

\newsec{Closed Amplitudes}

For closed amplitudes, we have a similar situation. A generic ansatz for 4-point functions is 
\eqn\clopA{
{\cal N}^{closed}_4 = \alpha^{(IJ) (KL)} {\Omega_I^{(1|1)} \Omega_J^{(1|1)}  {\cal K}_{KL}^{(1|1)} + 
\beta^{IJ KL} {\cal J}^{(1|2)}_{IJ} \overline{\cal J}^{(2|1)}_{KL}  }\,. 
}
The BRST closure yields two solutions: 
$\beta^{IJKL} = \alpha^{(IJ) (KL)}$, with a generic $\alpha^{(IJ) (KL)}$, and $\alpha^{(IJ)(KL)}=0$ with $\beta^{IJ KL} = \beta^{[IJ] [KL]}$. In the first case we can rewrite the amplitude as a 
trivial expression as 
\eqn\clopB{
{\cal N}^{clos}_4 = - {1\over 2} Q_L Q_R \Big( \alpha^{(IJ) (KL)} {{\cal K}^{(1|1)}_{IJ} {\cal K}_{KL}^{(1|1)} }
 \Big)\,,
}
which drops out of the computation, while the second case gives the following non-trivial expression  
\eqn\CloseC{
{\cal N}^{closed}_4\Big(ij,uv,kl,wz \Big) = 
{ {\cal J}^{(1|2)}_{ij|uv}  \overline{\cal J}^{(2|1)}_{kl|wz} \over s_{ij} } +  
{ {\cal J}^{(1|2)}_{ij|uw}  \overline{\cal J}^{(2|1)}_{kl|vz} \over s_{ij} } +  
{ {\cal J}^{(1|2)}_{ik|uv}  \overline{\cal J}^{(2|1)}_{jl|wz} \over s_{ik} } +  
{ {\cal J}^{(1|2)}_{ik|uw}  \overline{\cal J}^{(2|1)}_{jl|vz} \over s_{ik} }\,,
}
where we have reconverted the polarizations of the entering particles as pairs of indices 
$ij, kl,  uv, wz$. We have used the above relations to simplify the expression of the amplitude. 
It can be checked that acting with $Q_L$ and $Q_R$, respectively, 
${\cal N}^{clos}_4\Big(ij,uv,kl,wz \Big)$ is invariant and it cannot be written as a BRST variation. 

For 5-point function, we have the following ansatz
\eqn\closedfiveA{
{\cal N}^{closed}_5= 
\alpha^{IJK (LM)} {\cal K}^{(1|1)}_{IJK} \Omega^{(1|1)}_L  \Omega^{(1|1)}_M + 
\beta^{((IJ)(KL))M} {\cal K}^{(1|1)}_{IJ} {\cal K}^{(1|1)}_{KL}  \Omega^{(1|1)}_M +
}
$$
+\gamma^{IJK (LM)} \overline{\cal J}^{(2|1)}_{IJK} {\cal J}^{(1|2)}_{LM} + 
\delta^{IJK (LM)} {\cal Y}^{(1|2)}_{IJK} \bar{\cal J}^{(2|1)}_{LM} + 
\epsilon^{IJK (LM)} {\Delta}^{(2|2)}_{IJK} {\cal K}^{(1|1)}_{LM}\,. 
 $$
Imposing $Q_L$ and $Q_R$ invariance we get $\alpha=\gamma, \beta=0$ and $\delta = - \epsilon$, and $\alpha=- \delta, \beta=0$ and $\gamma = - \epsilon$ respectively. These equations have two different solutions: on one hand $\alpha=\gamma=-\delta=\epsilon$ and $\beta=0$; on the other hand $\alpha=\gamma=\delta=\epsilon=0$ and $\beta^{((IJ)(KL))M} + \beta^{((IJ)(ML))K}=0$. The first solution  implies that the final form of the amplitude can be written as
\eqn\closedfiveB{
{\cal N}^{closed}_5= 
\alpha^{IJK (LM)} \Big(
{\cal K}^{(1|1)}_{IJK} \Omega^{(1|1)}_L  \Omega^{(1|1)}_M +
 {\cal J}^{(2|1)}_{IJK} {\cal J}^{(1|2)}_{LM} - 
 {\cal Y}^{(1|2)}_{IJK} \bar{\cal J}^{(2|1)}_{LM} + 
 {\Delta}^{(2|2)}_{IJK} {\cal K}^{(1|1)}_{LM}\Big)\,, 
}
and, in turn, it can also be written as a BRST exact expression of the 
form 
\eqn\closedfiveC{
{\cal N}^{closed}_5= 
Q_R Q_L \Big(\alpha^{IJK (LM)} 
{\cal K}^{(1|1)}_{IJK} {\cal K}^{(1|1)}_{LM}\Big)\,.
}
Again, summing up all possible terms, the amplitude is 
BRST exact. This tells us that we can extract a BRST invariant amplitude by 
choosing a single term in \closedfiveC\ selecting for example the coefficient to be 
$\alpha^{IJK(LM)}(a,b,c,d,e) = \delta^I_a \delta^J_b \delta^K_c \delta^{(L}_d \delta^{M)}_e$. 
However, the second solution, namely $\beta^{((IJ)(KL))M} + \beta^{((IJ)(ML))K}=0$ leads to 
the physical amplitude. In the following section, we will provide the complete expression 
written in terms of nicer invariants. 

Remarks: notice that the expression for the closed amplitude as in \closedfiveC\ reminds the expression for the open 5-point 
amplitudes \openfiveC\ where the single BRST charge is replaced by the charges $Q_L \otimes Q_R$ and 
the building blocks of the open strings are replaced by the corresponding closed string ones 
$$
{\cal M}_{ijk} \rightarrow {\cal K}_{IJK}\,, \quad\quad 
{\cal M}_{ij} \rightarrow {\cal K}_{IJ}\,. 
$$


\subsec{4-point Closed String Amplitude}

In this section we build 4-point closed string amplitude using the method 
explained in \BernIA, which we briefly recall: any scattering amplitude for open string can be decomposed as a sum over graphs (labeled by index $\a$):
$$
{\cal{A}}^{open}_m = g^{m-2} \sum_\a {{c_\a n_\a}\over{\prod_{i} p^2_{\alpha_j}}}\,,
$$
where $c_\a$ is the color factor and $n_\a$ is the kinematic factors for the graph $\a$. The BCJ duality proposes that, for any set of three graphs $\a_1$, $\a_2$ and $\a_3$ related by a color Jacobi identity, there is a corresponding relation involving only the kinematic factors. Inspired by this fact, one can decompose the amplitudes in a dual way:
$${\cal{A}}^{open}_m = g^{m-2} \sum_\sigma \tau(\sigma) A^{dual}(\sigma)\,,$$
where $\sigma$ labels all non-cyclic permutations of the external legs and $A^{dual}(\sigma)$ denotes the amplitude decomposed in terms of kinematical factors 
and not in terms of color factors. Assuming BCJ duality to be true, a proposal for the closed string amplitudes is \BernIA:
\eqn\csa{{\cal{A}}^{cloesd}_m = i \left( {\kappa\over 2} \right)^{m-2} \sum_\sigma \tau^L(\sigma) {\cal{A}}^R_m(\sigma)\,. }

In order to show how this proposal works, let us evaluate ${\cal{A}}^{closed}_4$. First of all, looking at the open string amplitude ${\cal{A}}_4^{open}$, we have to identify the $n_{ij(kl)}$ kinematic terms. This is an easy task, since 
\eqn\OSAfour{
{\cal{A}}^{open}_4 (1234) \equiv {{\langle{V}_1 {V}_2 T_{34}\rangle} \over {s_{12}}} + {{\langle{V}_2 {V}_3 T_{41}\rangle} \over {s_{14}}} = {n_{12(34)}\over{s_{12}}} + {n_{23(41)} \over {s_{14}}}\,.
}
The $n$'s obey the Jacobi-like identity
$$n_{12(34)} - n_{23(41)} - n_{42(31)} = 0\,,$$
which can be deduced from the fact that the combination of the following kinematic terms
\eqn\JL{
{V}_1 {V}_2 T_{34} - {V}_2 {V}_3 T_{41} - {V}_4 {V}_2 T_{31} = {V}_1 {V}_{\{2} T_{34\}} + {Q} \left( {T_{23} T_{14} \over s_{14}} + {T_{24} T_{31} \over s_{13}}\right) = }
$$= {Q} \left( {T_{23} T_{14} \over s_{14}} + {T_{24} T_{31} \over s_{13}} + {{V}_1 T_{234} \over {s_{14}}} \right)\,,$$
is ${Q}$-exact.
The kinematic numerators $n_j$ are needed in order to obtain the building blocks of the closed string amplitude, the kinematic dual terms $\tau_j$ which are given by the definition
\eqn\taun{n_{ij(kl)} = \tau_{(i[j[k,l]])} \equiv \tau_{(ijkl)} - \tau_{(ijlk)} - \tau_{(iklj)} + \tau_{(ilkj)}\,.}
The $\tau$'s enjoy a number of properties: first of all, the same cyclic property of the color traces ($\tau_{(ijkl)} = \tau_{(jkli)} = \tau_{(klij)} = \tau_{(lijk)} $). Second, some of them are related by the Kleiss-Kuijf identities \KleissNE\ 
\eqn\kk{\tau_{1\{\alpha\}p\{\beta\}} = (-1)^{|\beta|} \sum_{\{\sigma\}} \tau_{1\{\sigma\}p}\,,} 
where $\{\alpha\}$ and $\{\beta\}$ are any subset of indices, $|\beta|$ is the number of elements of the set $\{\beta\}$, $\{\sigma\}$ is an element of the set of ordered permutations $OP(\{\alpha\},\{\beta^T \})$. Using the above relations, it is possible to show that, in the case of 4-point functions, only 2 different $\tau$'s are independent, and we choose them to be $\tau_a = \tau_{(1234)}$ and $\tau_b = \tau_{(1342)}$. Eq. \taun\ can obviously be inverted, to give the $\tau$'s in terms of the $n$'s:
$$\tau_{(ijkl)} = {1\over 6} \left( n_{ij(kl)} + n_{jk(li)} \right)\,.$$
Equipped with the dual kinematic terms, we can now rewrite the amplitudes as
$${\cal{A}}^{open}_4(ijkl) = \left({2\over{s_{ij}}} + {2\over{s_{il}}} \right) \tau_{(ijkl)} - \left( {2\over{s_{ij}}} \right) \tau_{(ijlk)} - \left( {2\over{s_{il}}} \right) \tau_{(iljk)} \,,$$
and the $4$-point closed string amplitude, after some manipulation, turns out to be ($\sigma$ now labels all the in-equivalent permutations)
\eqn\cloamp{{\cal{A}}^{closed}_4 \equiv i \left( {\kappa\over 2} \right)^2 \sum_\sigma \tau^L(\sigma) {\cal{A}}^R_m(\sigma) =} 
$$ = 4 i \left( {\kappa\over 2} \right)^2 \left[ {{(\tau^L_a - \tau^L_b ) (\tau^R_a - \tau^R_b)}\over{s_{12}}} + {{(\tau^L_a + 2 \tau^L_b ) (\tau^R_a + 2 \tau^R_b)}\over{s_{13}}} + {{(2 \tau^L_a + \tau^L_b ) (2 \tau^R_a + \tau^R_b)}\over{s_{14}}} \right]\,.$$

The ${Q}$-closure of the above amplitude can be deduced from the fact that, by construction, 
$${\cal{A}}^{closed}_4 \equiv i \left( {\kappa\over 2} \right)^2 \sum_\sigma \tau^L(\sigma) {\cal{A}}^R_m(\sigma) = i \left( {\kappa\over 2} \right)^2 \sum_\sigma {\cal{A}}^L_m(\sigma) \tau^R(\sigma) \,$$
and, while the first expression is clearly closed under the action of ${Q}^R$, the second is closed under ${Q}^L$, so that the whole amplitude is closed under the action of ${Q} = {Q}^L + {Q}^R$.


\subsec{5-point Supergravity Amplitudes}

In analogy to what has been done in the previous subsection, we now construct $5$-points closed string amplitude ${\cal{A}}^{closed}_5$. The $n$'s can be guessed by looking at the open string amplitude given, for instance, in \MafraIR
\eqn\OSAfive{
{\cal{A}}^{open}_5 (12345) \equiv {{\langle T_{12} {V}_3 T_{45} \rangle} \over {s_{12} s_{45}}} + {{\langle T_{23} {V}_4 T_{51} \rangle} \over {s_{23} s_{15}}} + {{\langle T_{34} {V}_5 T_{12} \rangle} \over {s_{12} s_{34}}} + {{\langle T_{45} {V}_1 T_{23} \rangle} \over {s_{23} s_{45}}} + {{\langle T_{51} {V}_2 T_{34} \rangle} \over {s_{34} s_{15}}} = }
$$= {n_{12(3(45))}\over{s_{12} s_{45}}} + {n_{23(4(51))}\over{s_{23} s_{15}}} + {n_{34(5(12))}\over{s_{12} s_{34}}} + {n_{45(1(23))}\over{s_{23} s_{45}}} + {n_{51(2(34))}\over{s_{34} s_{15}}}, $$
that is $n_{ij(k(lm))} \equiv \langle T_{ij} {V}_k T_{lm} \rangle$.
The kinematic dual terms $\tau$ are defined as in \taun\ to be
\eqn\taunfive{n_{ij(k(lm))} = \tau_{(i[j[k[l,m]]])}.}
Again, not all the $\tau$'s are independent, since they are constrained by Kleiss-Kuijf relations. There are two classes of KK relations, depending on the choice of the ``pivot'' index $p$ of formula \kk
$$\tau_{1ijkl} = \tau_{1ilkj} + \tau_{1likj} + \tau_{1lkij}\,,$$
$$\tau_{1ijkl} = - \tau_{1ijlk} - \tau_{1iljk} - \tau_{1lijk}\,,$$
for any choice of $\{i,j,k,l\}$. Using those relations it is possible to show that only 6 $\tau$'s out of 24 are independent, and that the following equation, relating back the $\tau$'s with the $n$'s is true
$$\tau_{(ijklm)} = {1\over{20}} \left( n_{ij(k(lm))} + n_{jk(l(mi))} + n_{kl(m(ij))} + n_{lm(i(jk))} + n_{mi(j(kl))} \right).$$ 

In order to write up the amplitudes it is convenient to set up the following notation: we denote by $f_a(s)$ with $a=1,\dots,15$ the 
quadratic combinations of momenta of the different channels in the 5-point amplitudes. They are algebraically related to the six independent momenta needed 
to parametrize the amplitudes completely. In addition, we introduce the partial amplitudes $\omega^L_a$ and $\omega^R_a$ 
defined by the following linear combinations 

\eqn\partamp{
\omega^L_1 = \left( \tau^L_{(1 5 2 4 3)} + \tau^L_{(1 5 3 2 4)} + 2\,\left( \tau^L_{(1 5 3 4 2)} + \tau^L_{(1 5 4 2 3)} + 2\,\tau^L_{(1 5 4 3 2)} \right)  \right) \,,
}
$$
\omega^L_2=   \left( \tau^L_{(1 5 2 3 4)} + 2\,\tau^L_{(1 5 3 2 4)} + 4\,\tau^L_{(1 5 3 4 2)} + \tau^L_{(1 5 4 2 3)} + 2\,\tau^L_{(1 5 4 3 2)} \right) \,,
$$
$$
\omega^L_3= \left( \tau^L_{(1 5 2 3 4)} - \tau^L_{(1 5 2 4 3)} + \tau^L_{(1 5 3 2 4)} + 2\,\tau^L_{(1 5 3 4 2)} - \tau^L_{(1 5 4 2 3)} - 2\,\tau^L_{(1 5 4 3 2)} \right)\,,
$$
$$
\omega^L_4= \left( 2\,\tau^L_{(1 5 2 3 4)} + \tau^L_{(1 5 2 4 3)} - 2\,\tau^L_{(1 5 3 2 4)} - \tau^L_{(1 5 3 4 2)} + \tau^L_{(1 5 4 2 3)} - 
     \tau^L_{(1 5 4 3 2)} \right) \,,
$$
$$
\omega^L_5=\left( 4\,\tau^L_{(1 5 2 3 4)} + 2\,\tau^L_{(1 5 2 4 3)} + 2\,\tau^L_{(1 5 3 2 4)} + 
     \tau^L_{(1 5 3 4 2)} + \tau^L_{(1 5 4 2 3)} \right)\,,
$$
$$
\omega^L_6= \left( 2\,\tau^L_{(1 5 2 3 4)} + \tau^L_{(1 5 2 4 3)} + 4\,\tau^L_{(1 5 3 2 4)} + 
     2\,\tau^L_{(1 5 3 4 2)} + \tau^L_{(1 5 4 3 2)} \right)\,,
$$
$$
\omega^L_7= \left( \tau^L_{(1 5 2 3 4)} + 2\,\tau^L_{(1 5 2 4 3)} + \tau^L_{(1 5 3 4 2)} + 4\,\tau^L_{(1 5 4 2 3)} + 2\,\tau^L_{(1 5 4 3 2)} \right)\,,
$$
$$
\omega^L_8= \left( \tau^L_{(1 5 2 3 4)} + 2\,\tau^L_{(1 5 2 4 3)} + \tau^L_{(1 5 3 2 4)} - \tau^L_{(1 5 3 4 2)} - 2\,\tau^L_{(1 5 4 2 3)} - \tau^L_{(1 5 4 3 2)} \right)\,,
$$
$$
\omega^L_9=\left( 2\,\tau^L_{(1 5 2 3 4)} + 4\,\tau^L_{(1 5 2 4 3)} + \tau^L_{(1 5 3 2 4)} + 2\,\tau^L_{(1 5 4 2 3)} + \tau^L_{(1 5 4 3 2)} \right)\,,
$$
$$
\omega^L_{10}=\left( \tau^L_{(1 5 2 3 4)} - \tau^L_{(1 5 3 2 4)} - \tau^L_{(1 5 4 2 3)} + \tau^L_{(1 5 4 3 2)} \right)\,,
$$
$$
\omega^L_{11}= \left( \tau^L_{(1 5 2 4 3)} - \tau^L_{(1 5 3 2 4)} + \tau^L_{(1 5 3 4 2)} - \tau^L_{(1 5 4 2 3)} \right)\,,
$$
$$
\omega^L_{12}=\left( \tau^L_{(1 5 2 3 4)} - \tau^L_{(1 5 2 4 3)} - \tau^L_{(1 5 3 4 2)} + \tau^L_{(1 5 4 3 2)} \right)\,,
$$
$$
\omega^L_{13}=\left( \tau^L_{(1 5 2 3 4)} + \tau^L_{(1 5 2 4 3)} - \tau^L_{(1 5 3 2 4)} - \tau^L_{(1 5 3 4 2)} + 2\,\tau^L_{(1 5 4 2 3)} - 2\,\tau^L_{(1 5 4 3 2)} \right)\,,
$$
$$
\omega^L_{14}= \left( \tau^L_{(1 5 2 3 4)} + \tau^L_{(1 5 2 4 3)} + 2\,\tau^L_{(1 5 3 2 4)} - 2\,\tau^L_{(1 5 3 4 2)} - \tau^L_{(1 5 4 2 3)} - 
     \tau^L_{(1 5 4 3 2)} \right)\,,
$$
$$
\omega^L_{15}=\left( 2\,\tau^L_{(1 5 2 3 4)} - 2\,\tau^L_{(1 5 2 4 3)} + \tau^L_{(1 5 3 2 4)} + 
     \tau^L_{(1 5 3 4 2)} - \tau^L_{(1 5 4 2 3)} - \tau^L_{(1 5 4 3 2)} \right)\,.
$$
Analogously for the right movers. The above expressions are not independent, but they are function 
of the six independent amplitudes $\tau^{L/R}_{(1 5 2 3 4)}, \dots,\tau^{L/R}_{(1 5 4 3 2)}$. 

It is now possible to make use of \csa\ to write the closed string amplitudes in terms of the (left and right) dual kinematic terms
as follows
\eqn\clofive{{\cal{A}}^{closed}_5 = \sum_{a=1}^{15} {\omega^L_a \omega^R_a \over f_a(s)} \,,}
showing the desired factorization properties in terms of Mafra {\it et al.}'s building blocks 
for open string theory. The above expression is adapted for 10d supergravity and therefore the expansion 
of the superfields in the blocks would contains also the dependence on the RR fields of the supergravity 
spectrum. 

Moreover, \clofive\ is ${Q}$-closed, as can be shown upon use of 
\eqn\qtau{ {Q} T_{ijklm} = 
(s_{ij} - s_{jk}) {V}_i {V}_j {V}_k T_{lm} 
+ (s_{jk} - s_{kl}) {V}_j {V}_k {V}_l T_{mi} 
+ (s_{kl} - s_{lm}) {V}_k {V}_l {V}_m T_{ij}}
$$+ (s_{lm} - s_{mi}) {V}_l {V}_m {V}_i T_{jk} 
+ (s_{mi} - s_{ij}) {V}_m {V}_i {V}_j T_{kl}\,.$$
where $\tau_{ijklm} = \langle T_{ijklm}\rangle$. 

\newsec{Conclusions}

Here, we take the first steps towards a complete analysis of supergravity amplitudes using pure spinor building blocks as inspired by Mafra's work. We construct the closed string theory building blocks and we compute the amplitudes in terms of those. We discuss the factorization properties. 
Nevertheless, there are several open issues than can be tackled in the present framework, but 
it is interesting to see how well these structures fit into the decomposition of the amplitudes taking into 
account the supersymmetric invariance of the underlying string theory model 
(a very similar analysis has been performed by the authors of 
\BianchiPU). Higher point and 
higher loop amplitudes will be discussed elsewhere. 

\newsec{Acknowledgments}

We are grateful to Pierre Vanhove for useful discussions. P.A.G. is grateful for the hospitality at {\it iPhT} at Saclay where a preliminary part of the present work has been done.


\newsec{Appendix: Closed BCJ Relations}

To derive the new BCJ relation for closed string, we first observe 
that for 3-point function the following Maldestam relations are valid
\eqn\ClosedF{
(s_{uv}+s_{vw}+s_{wu})=0\,, \quad
(s_{ij}+s_{jk}+s_{ki}) =0\,,
}
where we have denoted a legs of a 3-point function with the pairs $iu, jv, kw$, and 
$s_{ij}=s_{uv}\,, s_{jk}=s_{vw}, s_{ki}=s_{wu}$.  
Then we multiply the two Maldestam relations and their product with the product 
of $\Omega^{(1,1)}_{iu} \,, \Omega^{(1,1)}_{jv}\,, \Omega^{(1,1)}_{kw}$ to get 
$$
(s_{uv}+s_{vw}+s_{wu})(s_{ij}+s_{jk}+s_{ki}) \Omega^{(1,1)}_{iu} \, 
\Omega^{(1,1)}_{jv}\, \Omega^{(1,1)}_{kw} =0\,.
$$
Now, we decompose it into the different pieces obtained by summing cyclically on each triplet 
$ijk$ and $uvw$. Each term can be rewritten as follows 
$$
s_{uv} s_{ij} \Omega^{(1,1)}_{iu} \, \Omega^{(1,1)}_{jv}\, \Omega^{(1,1)}_{kw} = 
s_{uv} Q_L 
\Big({\cal J}^{(1,2)}_{ij|uv} \Omega^{(1,1)}_{kw}\Big) = - Q_L Q_R 
\Big({\cal K}^{(1,1)}_{ij|uv} \Omega^{(1,1)}_{kw}\Big)\,.
$$
So, finally we can conclude that the generalized BCJ relations 
can be written as
$$
{\cal K}^{(1,1)}_{\{ij|\{uv} \Omega^{(1,1)}_{k\}w\}} = Q_L Q_R {\cal W}^{(1,1)}_{ijk|uvw}\,.
$$

\listrefs
\bye